# Toward a Unified Performance and Power Consumption NAND Flash Memory Model of Embedded and Solid State Secondary Storage Systems

Pierre Olivier*+, Jalil Boukhobza*, Eric Senn+
Université Européenne de Bretagne, CNRS, UMR 6285 Lab-STICC, France
*Université de Bretagne Occidentale, Brest, France
+Université de Bretagne Sud, Lorient, France
*{firstname.lastname}@univ-brest.fr, +{firstname.lastname}@univ-ubs.fr

*Abstract*—This paper presents a set of models dedicated to describe a flash storage subsystem structure, functions, performance and power consumption behaviors. These models cover a large range of today's NAND flash memory applications. They are designed to be implemented in simulation tools allowing to estimate and compare performance and power consumption of I/O requests on flash memory based storage systems. Such tools can also help in designing and validating new flash storage systems and management mechanisms. This work is integrated in a global project aiming to build a framework simulating complex flash storage hierarchies for performance and power consumption analysis. This tool will be highly configurable and modular with various levels of usage complexity according to the required aim: from a software user point of view for simulating storage systems, to a developer point of view for designing, testing and validating new flash storage management systems.

## I. INTRODUCTION

Flash memory is the main secondary storage media in embedded systems due to its numerous benefits in terms of performance, shock resistance, data density and power consumption. It is also widely present in general computing and mass storage domains with the large adoption of flash based Solid State Drives (SSDs).

Simulation tools for such storage systems implement models of flash systems structure and behaviors. They are essential for multiple reasons. They first allow to evaluate and compare various performance metrics for existing systems. They also help in prototyping and validating new flash storage systems and the related management mechanisms. In the latter case, these tools give valuable estimations for performance and power consumption during the design phase. The aim of the work presented in this paper is to define a flash memory model which will be implemented in a global flash based storage system simulation tool. It will allow the simulation of a large set of storage systems from single flash chip embedded storage to complex multi chip and highly parallel SSDs.

Existing tools [1-4] allow simulating flash based systems to study their behaviors. The targeted level of description goes from the micro architectural level [1] to the system (SSD) level [2-4]. As they are useful for the reasons stated above, some of them provide simple models for flash systems that could be enhanced to describe today's complex storage systems, while others target low level descriptions which is too detailed for system level simulation. Most of them target I/O performance and could be completed with power consumption metrics. Another common characteristic is that these tools use constants as input for latency and power consumption profiles. This is somewhat rigid and could be enhanced by introducing models for describing performance and power consumption behaviors. Finally, such tools may benefit from detailed documentation to reduce the learning curve for mastering the tool.

In this paper we present a set of models designed to be implemented in simulation and estimation tools, targeting NAND flash based storage systems performance and power consumption. The models are designed in such a way that covers the large specter of today's NAND flash applications. With these models one can describe simple systems like single NAND chip based storage systems for embedded boards (smartphones, tablet PCs, etc.). One can also describe complex multi-chip, multi-channels systems as those present in SSDs. The presented models allow the user to define the structural and functional parameters for the described flash subsystem. Performance and power consumption models are defined, allowing to compute execution time and energy consumed by the various events. An implementation of these models in a fully documented C++ library is proposed in this paper. It will be available along with several use cases and performance / power consumption models derived from real NAND based storage systems. This tool takes as input a description of the simulated system and an I/O trace.

In a first section NAND flash memory is described as well as the different ways to manage it in computer systems. In a second section the models for NAND flash storage systems are described. In section 3 an implementation example of the models is given. Finally, the global project in which this work is integrated is depicted in section 4, and some conclusions and perspectives of future works are given.

## II. BACKGROUND ON NAND FLASH MEMORY

### A. Structure of a NAND Flash Memory Based Storage System

Flash memory is a non volatile EEPROM memory. NAND flash subtype [5] is block addressed and offers a high storage density: it is used for data storage. The structure of a complex NAND storage subsystem [4] is depicted on Fig. 1. A chip is composed of one or more dies, each of them containing one or more planes. Planes contain blocks, themselves containing pages. A page is composed of a user data area for data storage, and an *out-of-band* area to store metadata. A plane contains a page buffer with the size of one page to buffer data read / written from / to the chip. As simple embedded systems usually use only one NAND chip for secondary storage, more complex systems like SSDs contain multiple chips. Chips sharing the same I/O bus are grouped into channels.

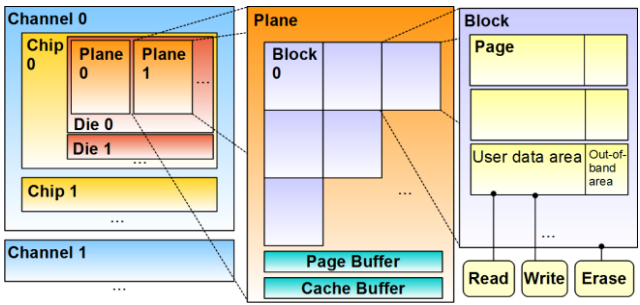

Fig. 1. A complex NAND storage subsystem structure.

*B. NAND Legacy and Advanced Commands*

A NAND flash storage subsystem, regardless of its structure complexity, supports 3 main operations called legacy operations: *read* and *write* operations are achieved at the granularity of a page, and the *erase* operation is performed on a whole block (see Fig. 1.).

More complex systems like SSDs make use of advanced NAND commands [4]: (1) the *copy-back* command allows to use the page buffer to move data from one page to another inside one plane. (2) The *cache read* and *cache write* commands allows using the cache buffer to pipeline data transfers between the NAND array and the I/O bus (see Fig. 1.), enhancing data transfer speed. The following commands introduce the notion of parallelism for I/O processing. (3) *Multi-plane read / write / erase* launch in parallel several read / write / erase operations in all the planes of a die. (4) *Die interleaved read / write / erase* do likewise for several dies in the same chip. Finally, commands can be launched in parallel in multiple chips of the same channels, and in multiple channels of the storage system. Advanced operations can also be combined, for example a multi plane copy back operation can be performed in one die.

*C. NAND Constraints and Management Mechanisms*

NAND flash exhibits specific constraints. The first is the *erase-before-write* rule, which state that a page must first be erased before being written. As the target of the erase operation is an entire block, flash management mechanisms cope with this constraint by performing out-of-place data updates. This implies the use of a logical to physical address mapping scheme, and the invalidation of old (updated) data versions. Invalid data are recycled asynchronously through the execution of a process called garbage collection. Another constraint is the fact that a flash memory block can only sustain a limited number of erase operations, after which it can no longer retain data. Management mechanisms must then distribute evenly the write and erase cycles over the whole flash array to maximize the flash lifetime.

Some constraints also apply specifically on advanced operations: for example, due to the internal characteristics of a flash chip, the addresses of source and target pages of a copy back operation must be either both odd or both even.

III. MODELING A NAND FLASH STORAGE SUBSYSTEM

We describe a flash storage subsystem as a set of 4 models: the structural, fuctional, performance, and power model.

The *structural model* describes the architecture of the NAND subsystem : how it is divided into channels, chips, dies, planes, blocks and pages. It also allows to specify the architectural parameters of the represented flash subsystem: for example the page size, number of pages per block, number of channels, etc. With such a structural model one can choose to describe a single chip as an example of simple embedded flash storage system or a complex SSD storage subsystem.

The *functional model* is used to describe the I/O operations supported by the simulated flash subsystem: we model legacy and advanced operations. The way the system processes these operations is also modeled: for example a copy back operation will end up as a read operation on the source page, and a write operation on the target page. Another important objective of the functional model is to implement the various NAND constraints presented earlier. An example of input for the functional model can be the list of commands supported by the described system: legacy commands only for a simple embedded storage subsystem, or a full set of advanced commands for a complex SSD subsystem.

The *performance model* describes how to compute the execution time of the various events that can occur in the flash system. It can be viewed as a set of equations, one for each event, and the associated parameters. For example when the flash system processes a page read request, the performance model provides a function implementing an equation to compute the execution time of the page read operation. This function takes as input some meaningful parameters, for example the address of a page to read.

The *power consumption* model works the same way as the performance model. Each function of the power consumption model implements an equation for the related flash event. In addition to the parameters that are also provided to the performance model, the power consumption model takes also as input the latencies computed by the performance model, as the time is a key metric in energy calculation.

The power consumption and performance models abovementioned can be seen as meta-models. Indeed, they describe the parameters that can be used to compute execution time or power consumption for various events occurring in the described flash memory system. For each of these events, the equations using these parameters can be provided by the user. These equations are the actual performance and power consumption models. Such a high level of abstraction allows an accurate description of various systems with strong specificities in performance / power consumption behaviors.

## IV. IMPLEMENTATION: A C++ API FOR FLASH SUBSYSTEM PERFORMANCE / POWER CONSUMPTION ESTIMATION

The structural model (A on Fig. 2) is defined as a set of C++ classes, one for each subcomponent of the flash subsystem (see Fig. 1.). When instantiating a flash subsystem object, structural parameters are passed to the constructor.

The functional model corresponds to the methods associated to these classes. The "top level" component representing the flash subsystem offers one interface for each of the supported flash operations. This component takes as input an I/O trace (B on Fig. 2) during the simulation. This trace is a list of flash commands with the related parameters (arrival time, address, etc.). I/O requests are received by the top level component which forwards them to the concerned subcomponents. As said earlier, the functional model also implements the NAND flash constraints. This is useful when designing new flash management systems to detect invalid operations then raise error or warning messages. Still with this aim in mind, each subcomponent performs address range checking on received I/O requests.

The functional model implementation computes the various flash events occurring during the simulation and passes them to the performance model for latency calculation (D, E). Along with the type of the event, meaningful parameters needed for timing computation are also passed: for example the address of a flash page read. In a same way the power consumption model receives events and parameters for energy computation (D, F, G). The outputs of performance and energy computation are passed to a statistics module (H, I), along with a log of occurred flash events (C). This module is in charge of preparing the simulation output.

A simulation is highly configurable: an I/O trace and architectural parameters for the structural model can be provided. At this point in time, the API is still under development. It will be able to receive custom performance and power consumption models as input. They will take the form of equations computing timings and energy for each flash event, based on the meaningful parameters. The implementation will be available online along with a set of pre-built models, in particular performance and power consumption models created from real flash storage systems.

## V. GLOBAL PROJECT: A NAND FLASH STORAGE SYSTEM SIMULATOR

This work is achieved for inclusion in a global project aiming to model and simulate the performance and power consumption of complex flash based devices. This includes on one hand embedded Linux storage systems. In that case the flash subsystem will receive an I/O trace computed by a functional model for the entire Linux flash storage software stack: the virtual file system, the dedicated flash file system, and the NAND driver. Models for each of these layers will be built, as well as performance and power consumption for the involved hardware components: CPU and RAM. On the other hand this project includes SSD and similar devices simulation. The flash subsystem will then take its I/O input trace from a functional model of a complex SSD controller. Performance and power consumption models for the controller and potential embedded cache memories (RAM / SRAM) found in such devices will be included in the simulator.

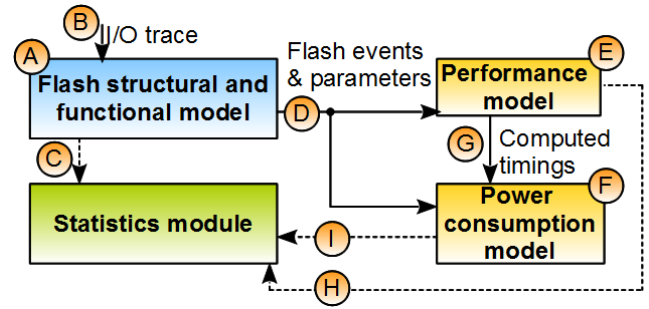

Fig. 2. Interactions between the models implementation inside the proposed C++ simulation API

This tool will have several functions: from a user point of view, one will be able to simply input parameters describing a flash storage system and its behavior to perform a simulation. From a developer point of view, bindings will be available to implement new flash management mechanisms for testing and validating purposes.

## VI. CONCLUSION AND FUTURE WORKS

We have presented a set of models allowing to describe a NAND flash memory subsystem structure, behavior, performance and power consumption characteristics. These models are designed to be implemented in simulation software tools essential to (1) evaluate and compare existing systems; (2) prototype and validate flash management mechanisms; (3) build and validate performance and power consumption models. These models are implemented in a fully documented C++ API allowing to simulate flash subsystems. This tool is highly configurable: in addition to the flash structural parameters and the input I/O trace, the API will support plugging custom performance and power consumption models. This work will be integrated in a flash based storage system simulation tool supporting the wide range of NAND flash based devices: from embedded storage to complex and highly parallel solid state drives.